\documentclass[12pt]{article}
\usepackage{graphicx}
\usepackage{latexsym}
\usepackage{amsmath,amsfonts,amsthm,bm}

\newcommand{\CGf}{\ensuremath{\mathcal{C}}}
\newcommand{\PGf}{\ensuremath{\mathcal{P}}}
\newcommand{\RGf}{\ensuremath{\mathcal{R}}}
\newcommand{\ave}[1]{\langle #1 \rangle}

\begin{document}

\title{Enumeration of self-avoiding walks on the square lattice}

\author{Iwan Jensen  \\
\small
ARC Centre of Excellence for Mathematics and Statistics of Complex Systems, \\
\small
Department of Mathematics and Statistics, \\
\small
The University of Melbourne, Victoria 3010, Australia}

\date{\today}

\maketitle

\begin{abstract}
We describe a new algorithm for the enumeration of self-avoiding walks
on the square lattice. Using up to 128 processors on a HP Alpha server cluster 
we have enumerated the number of self-avoiding walks on the square lattice 
to length 71. Series for the metric properties of mean-square end-to-end distance, 
mean-square radius of gyration and mean-square distance of monomers from the 
end points have been derived to length 59. Analysis of the resulting series
yields accurate estimates of the critical exponents $\gamma$ and $\nu$ confirming
predictions of their exact values. Likewise we obtain accurate amplitude
estimates yielding precise values for certain universal amplitude combinations.
Finally we report on an analysis giving compelling evidence that the leading
non-analytic correction-to-scaling exponent $\Delta_1=3/2$.
\end{abstract}

\numberwithin{equation}{section}

\section{Introduction}

The self-avoiding walk (SAW) on regular lattices is one of the most important 
and classic combinatorial problems in statistical mechanics \cite{MSbook}. 
SAWs are often considered in the context of lattice models of polymers. 
The fundamental problem is the calculation (up to translation) of the number 
of SAWs, $c_n$, with $n$ steps. As most interesting combinatorial problems, 
SAWs have exponential growth, $c_n \sim A\mu^n n^{\gamma-1}$, where $\mu$
is the connective constant, $\gamma = 43/32$ is a (known) critical exponent 
\cite{Nienhuis82a,Nienhuis84a}, and $A$ is a critical amplitude. So one major 
problem is the calculation, or at least accurate estimation of, $\mu$ and 
$\gamma$ in order to check the theoretical prediction. A second major problem is 
the calculation of critical amplitudes, such as $A$, in order to test predictions for  
various universal amplitude combinations for two-dimensional SAWs \cite{CS89,CPS90,CG93}. 
This requires, in addition to the calculation of $c_n$, the calculation of
metric properties such as the end-to-end distance and the radius of gyration.
Furthermore the enumeration of SAWs have traditionally served as a benchmark for 
both computer performance and algorithm design.

An {\em $n$-step self-avoiding walk} $\bm{\omega}$ on a regular lattice is 
a sequence of {\em distinct} vertices $\omega_0, \omega_1,\ldots , \omega_n$ 
such that each vertex is a nearest neighbour of it predecessor. SAWs are
considered distinct up to translations of the starting point $\omega_0$.
We shall use the symbol $\bm{\Omega}_n$ to mean the set of all 
SAWs of length $n$.

In addition we also consider self-avoiding polygons (SAPs). A SAP can be 
viewed as a SAW whose end-points $\omega_0$ and $\omega_n$ are 
nearest-neighbors and which therefore can be connected to form a closed loop 
by the addition of a single step. Notice that there are $2(n+1)$ SAWs which 
give rise to a given $(n+1)$ step SAP. Each vertex of the SAP can be used as 
$\omega_0$ and we could walk clockwise or counter-clockwise around the perimeter 
of the SAP.

The enumeration of SAWs and SAPs has a long and glorious history, which for
the square lattice has recently been reviewed in \cite{GC01}. Suffice to 
say that early calculations were based on various direct counting algorithms
of exponential complexity, with computing time $T(n)$ growing asymptotically 
as $\lambda^n$, where $\lambda = \mu \sim 2.638$, the connective constant 
for SAWs. Enting \cite{IGE80e} was the first to produce a major breakthrough 
by applying transfer matrix (TM) methods to the enumeration of SAPs on finite 
lattices. This so called finite lattice method (FLM) led to a very significant 
reduction in complexity to $3^{n/4}$, so $\lambda = \sqrt[4]{3}=1.316\ldots$. 
More recently we \cite{JG99} refined the algorithm using the method of pruning 
and reduced the complexity to $1.2^n$.
The extension of the FLM to SAW enumeration had to wait until 1993 when
Conway, Enting and Guttmann \cite{CEG93} implemented an algorithm with 
complexity  $3^{n/4}$. The algorithm is difficult to implement and requires
large amounts of physical memory. However, the algorithm cannot be used
to calculate metric properties. In this paper we pursue a different FLM 
algorithm based on the same ideas used to improve the SAP algorithm.
It appears that this pruning algorithm has a computational complexity of 
$1.334^n$ very close to the CEG algorithm. So the CEG will ultimately beat 
the pruning algorithm for large enough $n$. For small $n$ the pruning 
algorithm actually uses significantly less memory as we shall show in 
Section~\ref{sec:comp}, and it can in addition be used to calculate metric 
properties. To our knowledge this is the first time TM methods has been used 
to calculate metric properties of SAWs.

The quantities we consider in this paper are.

\begin{itemize}
\item The number of SAWs of length $n$, believed to have the asymptotic 
behaviour
\begin{subequations}
\begin{equation}\label{eq:sawasymp}
c_n  =  A \mu^n n^{\gamma-1}[1+o(1)],
\end{equation}
where $\mu$ is the connective constant and $\gamma$ is a critical
exponent. We shall also study the associated generating function 
\begin{equation}\label{eq:sawgf}
\CGf (u) = \sum_{n=0}^{\infty} c_n u^n = A\Gamma(\gamma) (1-u\mu)^{-\gamma}[1+o(1)],
\end{equation}
\end{subequations}
so the generating function has a singularity at $u=u_c=1/\mu$.
\item The number of SAPs of length $n$, believed to grow asymptotically as
\begin{subequations}
\begin{equation}\label{eq:sapasymp}
p_n  =  B \mu^n n^{\alpha-3}[1+o(1)],
\end{equation}
where $\alpha$ is another critical exponent. In this case
the generating function behaves as
\begin{equation}\label{eq:sapgf}
\PGf (u) = \sum_{n=0}^{\infty} p_{n} u^n = B\Gamma(\alpha-2)((1-u\mu)^{2-\alpha}[1+o(1)].
\end{equation}
\end{subequations}
\item The mean-square end-to-end distance 
of $n$ step SAWs 
\begin{subequations}
\begin{equation}\label{eq:ee}
\ave{R^2_e}_n = \frac{1}{c_n} \sum_{\bm{\Omega}_n} (\omega_0 - \omega_n)^2 =
 C n^{2\nu}[1+o(1)],
\end{equation}
where $\nu$ is a new critical exponent. We also look at
the generating function
\begin{equation}\label{eq:eegf}
\RGf_e (u) = \sum_{n} c_n \ave{R^2_e}_n u^n = 
              AC\Gamma(\gamma+2\nu)(1-u\mu)^{-(\gamma+2\nu)}[1+o(1)].
\end{equation}
\end{subequations}
\item The mean-square radius of gyration of $n$ step SAWs
\begin{subequations}
\begin{equation}\label{eq:rg}
\ave{R^2_g}_n = \frac{1}{c_n} \sum_{\bm{\Omega}_n}\left [ \frac{1}{2(n+1)^2} 
\sum_{i,j=0}^n (\omega_i - \omega_j)^2 \right ]=
 D n^{2\nu}[1+o(1)],
\end{equation}
with the associated generating function
\begin{equation}\label{eq:rggf}
\RGf_g (u) = \sum_{n} (n+1)^2 c_n \ave{R^2_g}_n u^n \
           = AD \Gamma(\gamma+2\nu+2)(1-u\mu)^{-(\gamma+2\nu+2)}[1+o(1)],
\end{equation}
where the factors under the sum ensure that the coefficients are integer 
valued.
\end{subequations}
\item The mean-square distance of a monomer from the end-points of 
$n$ step SAWs  
\begin{subequations}
\begin{equation}\label{eq:md}
\ave{R^2_m}_n = \frac{1}{c_n} \sum_{\bm{\Omega}_n} \left [ \frac{1}{2(n+1)}
\sum_{i=0}^n \left [(\omega_0-\omega_j)^2+(\omega_n-\omega_j)^2 \right ] \right ]
 = E n^{2\nu}[1+o(1)],
\end{equation}
with the associated generating function
\begin{equation}\label{eq:mdgf}
\RGf_m (u) = \sum_{n} (n+1)c_n \ave{R^2_m}_n u^n 
           = AE \Gamma(\gamma+2\nu+1)(1-u\mu)^{-(\gamma+2\nu+1)}[1+o(1)].
\end{equation}
\end{subequations}
\end{itemize}

The critical exponents are believed to be universal in that they only depend
on the dimension of the underlying lattice. $\mu$ on the other hand is non-universal.
For SAWs in two dimensions the critical exponents $\gamma = 43/32$, 
$\alpha =1/2$ and $\nu = 3/4$ have been predicted exactly, though 
non-rigorously, using Coulomb-gas arguments \cite{Nienhuis82a,Nienhuis84a}. 

While the amplitudes are non-universal, there are many universal amplitude
ratios. Any ratio of the metric amplitudes, e.g. $D/C$ and $E/C$, is expected 
to be universal \cite{CS89}. Many other universal amplitude combinations in 
particular involving SAPs can be found in \cite{CG93,RGJ01}. 
Of particular interest is the linear combination
\cite{CS89,CPS90} (which we shall call the CSCPS relation)
\begin{equation} \label{eq:CSCPS}
 F \;\equiv\;
   \left( 2 +  \frac{y_t}{y_h} \right)  \frac{D}{C}
   \,-\, 2 \frac{E}{C} \,+\, \frac12,
\end{equation}
where $y_t = 1/\nu$ and $y_h = 1 + \gamma/(2\nu)$ are the thermal and magnetic 
renormalization-group eigenvalues, respectively, of the $n$-vector model at 
$n=0$. In two dimensions ($y_t = 4/3$ and $y_h = 91/48$, hence 
$2 + y_t/y_h = 246/91$) Cardy and Saleur \cite{CS89} (as corrected by 
Caracciolo, Pelissetto and Sokal \cite{CPS90}) have predicted, using 
conformal field theory, that $F = 0$. This conclusion has been confirmed by 
previous high-precision Monte Carlo work \cite{CPS90} as well as by series 
extrapolations \cite{GY90}.

Privman and Redner \cite{PR85b} proved that the combination $BC/\sigma a_0$ 
is universal. $\sigma$ is an integer constant such that $p_n$ is non-zero 
when $n$ is divisible by $\sigma$. So $\sigma=1$ for the triangular lattice 
and 2 for the square and honeycomb lattices. $a_0$ is the area per lattice 
site and  $a_0=1$ for the square lattice, $a_0=3\sqrt{3}/4$ for the honeycomb 
lattice, and $a_0=\sqrt{3}/2$ for the triangular lattice.

The asymptotic form (\ref{eq:sawasymp}) for $c_n$ only explicitly gives
the leading contribution. In general one would expect corrections to
scaling so 
\begin{equation}
c_n= A\mu^n n^{\gamma-1}\left [1 + \frac{a_1}{n}+\frac{a_2}{n^2}+\ldots
+ \frac{b_0}{n^{\Delta_1}}+\frac{b_1}{n^{\Delta_1+1}}+\frac{b_2}{n^{\Delta_1+2}}+\ldots
\right]
\end{equation}
In addition to ``analytic'' corrections to scaling of the form $a_k/n^k$,
there are ``non-analytic'' corrections to scaling of the form
$b_k/n^{\Delta_1+k}$, where the correction-to-scaling exponent $\Delta_1$ 
isn't an integer. In fact one would expect a whole sequence of
correction-to-scaling exponents $\Delta_1 < \Delta_2 \ldots$, which
are both universal and also independent of the observable, that is,
the same for $c_n$, $p_n$, and so on. 
Much effort has been devoted to determining the leading non-analytic 
correction-to-scaling exponent $\Delta_1$ for two-dimensional SAWs and SAPs.
At least two different theoretical predictions have been made for the exact 
value of this exponent: $\Delta_1 = 3/2$ based on Coulomb-gas arguments
\cite{Nienhuis82a,Nienhuis84a}, and $\Delta_1 = 11/16$ based on 
conformal-invariance methods \cite{Saleur87a}.

In a recent paper \cite{CGJPRS} we studied the amplitudes and the
correction-to-scaling exponents for two-dimensional SAWs,
using a combination of series-extrapolation and Monte Carlo methods.
We enumerated all self-avoiding walks up to 59 steps on the square lattice,
and up to 40 steps on the triangular lattice, measuring the
metric properties mentioned above, and then carried out a detailed
and careful analysis of the data in order to accurately estimate the
amplitudes and correction-to-scaling exponent. In this
paper we give a detailed account of the algorithm used to calculate
the square lattice series analysed in \cite{CGJPRS}, report on a further
extension of the SAW counts up to 71 steps, analyse the series and confirm 
to great accuracy the predicted exact values of the critical exponents, and
finally we briefly summarise the results of the analysis from
\cite{CGJPRS}.

\section{Enumeration of self-avoiding walks \label{sec:flm}}

The algorithm we use to enumerate SAWs on the square lattice builds on the 
pioneering work of Enting \cite{IGE80e} who enumerated square lattice 
self-avoiding polygons using the finite lattice method. More specifically 
our algorithm is based in large part on the one devised by Conway, Enting and 
Guttmann \cite{CEG93} for the enumeration of SAWs. The basic idea of the finite 
lattice method is to calculate partial generating functions for various properties 
of a given model on finite pieces, say $W \times L$ rectangles of the square 
lattice, and then reconstruct a series expansion for the infinite lattice
limit by combining the results from the finite pieces. The generating
function for any finite piece is calculated using transfer matrix (TM)
techniques. 

\subsection{Basic transfer matrix algorithm}

\begin{figure}
\begin{center}
\includegraphics{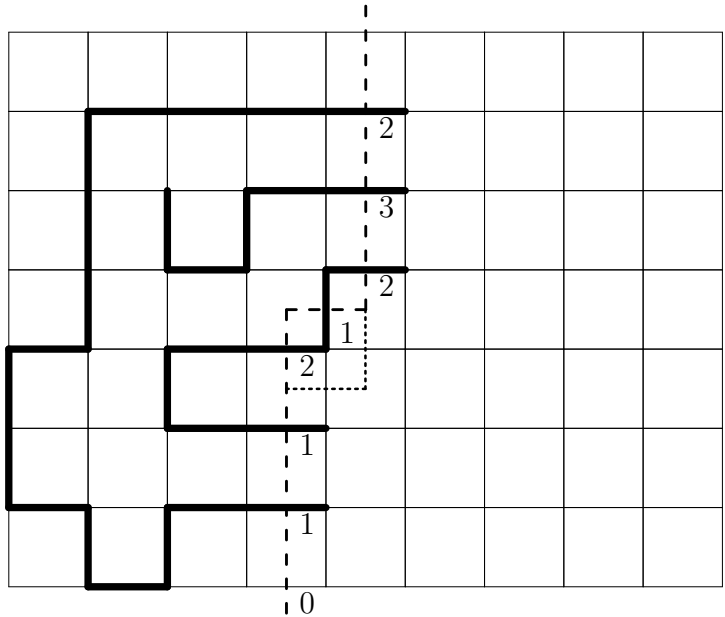}
\end{center}
\caption{\label{fig:transfer}
A snapshot of the boundary line (dashed line) during the transfer matrix 
calculation on the square lattice. SAWs are enumerated by successive
moves of the kink in the boundary line, as exemplified by the position given 
by the dotted line, so that one vertex at a time is added to the rectangle. 
To the left of the boundary line we have drawn an example of a 
partially completed SAW.}
\end{figure}

The most efficient implementation of the TM algorithm generally involves 
bisecting the finite lattice with a boundary (this is just a line in the 
case of rectangles) and moving the boundary in such a way as to build up 
the lattice cell by cell. The sum over all contributing graphs is 
calculated as the boundary is moved through the lattice. Due to the 
symmetry of the square lattice we need only consider rectangles with 
$L \geq W$. SAWs in a given rectangle are enumerated by moving the 
intersection so as to add  one vertex at a time, as shown in 
Fig.~\ref{fig:transfer}. For each configuration of occupied or empty edges 
along the intersection we maintain a generating function for partial walks 
cutting the intersection in that particular pattern. If we draw a SAW and 
then cut it by a line we observe that the partial SAW to the left of this 
line consists of a number of loops connecting two edges (we shall refer to 
these as loop ends) in the intersection, and pieces which are connected to 
only one edge (we call these free ends). The other end of the free piece is 
an end point of the SAW so there are at most two free ends. In applying the 
transfer matrix technique to the enumeration of SAWs we regard them as sets 
of edges on the finite lattice with the properties:
\begin{itemize}
\item[(1)] A weight $u$ is associated with an occupied edge. In some
cases one gives different weights $u$ and $v$ to occupied horizontal and 
vertical edges, respectively.
\item[(2)] All vertices are of degree 0, 1 or 2.
\item[(3)] There are at most two vertices of degree 1 and the final graph
has exactly two vertices of degree 1 (the end points of the SAW).
\item[(4)] Apart from isolated sites, the final graph has a single connected
component.
\item[(5)] In some implementations each graph must span the rectangle 
from left to right, while in other implementations each graph must span the
rectangle from left to right and from bottom to top.
\end{itemize}

We are not allowed to form closed loops, so two loop ends can only be joined 
if they belong to different loops. To exclude loops which close on themselves 
we need to label the occupied edges in such a way that we can easily determine 
whether or not two loop ends belong to the same loop. The most obvious choice 
would be to give each loop a unique label. However, on two-dimensional 
lattices there is a more compact scheme relying on the fact that two loops 
can never intertwine. Each end of a loop is assigned one of two labels 
depending on whether it is the lower end or the upper end of a loop. Each 
configuration along the boundary line can thus be represented by a set of 
edge states $\{\sigma_i\}$, where

\begin{equation}\label{eq:states}
\sigma_i  = \left\{ \begin{array}{rl}
0 &\;\;\; \mbox{empty edge},  \\ 
1 &\;\;\; \mbox{lower loop end}, \\
2 &\;\;\; \mbox{upper loop end}. \\
3 &\;\;\; \mbox{free end}. \\
\end{array} \right.
\end{equation}
\noindent
If we read from the bottom to the top, the configuration along the 
intersection of the partial SAW in Fig.~\ref{fig:transfer} is $\{011212320\}$. 
It is easy to see that this encoding uniquely describes 
which loop ends are connected. In order to find the upper loop end, matching a 
given lower end, we start at the lower end and work up wards  in the 
configuration counting the number of `1's and `2's we pass (the `1' of the 
initial lower end is {\em not} included in the count). We stop when the 
number of `2's  exceeds the number of  `1's. This `2'  marks the matching 
upper end of the loop. It is worth noting that there are some restrictions 
on the possible configurations. Firstly, every lower loop end must have a 
corresponding upper end, and it is therefore clear that the total number 
of `1's  is equal to the total number of `2's. Secondly, as we look through 
the configuration starting from the bottom the number of `1's is never 
smaller than the number of `2's. Ignoring the `0's the '1's and `2's can
be viewed as perfectly balanced parenthesis. Those familiar with algebraic 
languages will immediately recognize that each configuration (for now treating
free ends and empty edges in the same way) of labeled loop ends forms a Motzkin 
word \cite{DV84a}. 

\subsubsection{Derivation of updating rules}

The updating of a partial generating function depends on the states of the
edges to the left and above the new vertex. When the kink is moved
we insert the edges to the right and below  the new vertex. 
The way to avoid situations leading to graphs with more than a single
connected component is to forbid free ends from terminating (or joining)
at the vertex being processed unless the boundary line intersects no other 
occupied edges. In Table~\ref{tab:update} we have listed the possible local 
`input' states and the `output' states which arise as the kink in the boundary
is propagated by one step. The rows in this table are labeled by the state of
the left edge while the columns are labeled by the state of the top edge.
Each panel in the table contains the possible states of the bottom and 
right edges (in that order). We shall refer to the configuration before the move 
as the `source' and a configuration produced as a result of the move as a `target'.
In each move the source generating function is multiplied by $u^k$,
where $k$ is the number of new occupied edges (just the number of non-zero
entries in the local output state), and is then added to the target generating
function. 

In the following we give the details
of how some of these updating rules are derived. 

\begin{table}
\caption{\label{tab:update}
The various local `input' states and the `output' states which arise as the 
boundary line is moved in order to include one more vertex of the lattice.}
\begin{center}
\renewcommand{\arraystretch}{1.2}
\begin{tabular}{|c|cccc|ccc|ccc|ccc|}  \hline  \hline 
 &\multicolumn{4}{c|}{0} &\multicolumn{3}{c|}{1} & \multicolumn{3}{c|}{2}
& \multicolumn{3}{c|}{3}  \\ 
\hline
0  & $00$   & $12$   & $\bm{03}$   & $\bm{30}$ 
   & $01$   & $10$   & $\bm{\overline{00}}$
   & $02$   & $20$   & $\bm{\overline{00}}$
   & $03$   & $30$   & ADD \\    \hline
1  & $01$   & $10$   & $\bm{\overline{00}}$ & 
   &        & $\overline{00}$ &
   & \multicolumn{3}{c|}{NOT allowed}     
   &        &  $\overline{00}$ & \\  \hline
2  & $02$   & $20$   & $\bm{\overline{00}}$ & 
   &        & $00$   &
   &        &  $\overline{00}$ & 
   &        &  $\overline{00}$ & \\  \hline
3  & $03$   & $30$   & ADD & 
   &        & $\overline{00}$ &
   &        & $\overline{00}$ &
   &        &  ADD & \\
   \hline \hline
\end{tabular}
\end{center}
\end{table}

\begin{description}

\item{00:} The left and top edges are empty. We have four possible outputs. We can 
leave the bottom and right edges empty (00), insert a new partial loop (12), or add 
a new free end on the right (03) or bottom (30) edge. Adding a free end increases by 
one the number of degree-1 vertices, so this is only allowed provided the source has 
at most one free end. Throughout, this restriction is indicated by the use of boldface 
entries. 

\item{01,10,02,20:} The left or top edge is occupied by a loop end.
We can continue this loop end along either the right or bottom edge.
Note that we cannot occupy both new edges since this would lead to 
vertices of degree 3.
We can also leave both edges empty. This creates a new
degree-1 vertex and we have to relabel the matching end of the
discontinued loop as free. Relabeling is indicated by over-lining.
The way to identify the matching loop end is described below
(\ref{eq:states}) in the previous section.

\item{03,30:}  The left or top edge is occupied by a free end.
We can continue the free end along either the right or bottom edge.
We can also leave both edges empty. This creates a separate component and
is only allowed if the resulting graph is a valid SAW. That is,
the source contains no other occupied edges (and if required both
the bottom and top of the rectangle has been touched). The partial
generating function is added to the running total. We mark this
possibility by the entry ADD.

\item{11,22:} Two lower (upper) loop ends are joined and the output edges
must be empty (otherwise we would create vertices of degree greater than 2). 
The matching upper (lower) loop end of the inner-most loop 
is relabeled as the new lower (upper) end of the combined loop.

\item{12:} A closed loop would be formed. This is not allowed.

\item{21:}  Upper and lower loop ends are joined and the output edges
must be empty.

\item{13,31,23,32:} A free end is joined to a lower (upper) loop end. The
output edges are empty and the matching loop end is relabeled free.

\item{33:} Two free ends are joined. This results is a separate component.
If the resulting graph is a valid SAW we add it to the generating function.

\end{description}

\subsection{The Conway-Enting-Guttmann algorithm \label{sec:CEG}}

The algorithm used by Conway, Enting and Guttmann \cite{CEG93} 
to enumerate SAWs is ingenious but also quite complicated and relies heavily 
on manipulations of various generating functions. Here we shall only
give the briefest of outlines of the algorithm sufficient for the
reader to appreciate the differences between this algorithm and
the one we used for the enumerations reported in this paper.

The CEG algorithm leads to the enumeration of anisotropic SAWs,
that is the number of SAWs $c_{m,n}$ with $m$ steps parallel to
the $y$-axis and $n$-steps parallel to the $x$-axis. Obviously,
$c_{m,n}=c_{n,m}$. The major `trick' of the algorithm \cite{CEG93} is the
realisation that any SAW can be constructed by combining {\em irreducible} 
components. An irreducible component has at least two steps along the 
$y$-axis in each position, e.g., any line parallel to the $x$-axis will
intersect the component at least twice (or not at all if the line lies
beyond the extent of the component). There are 5 different types
of irreducible component. The irreducible components are 
obtained from enumerations of anisotropic SAWs in rectangles.
The SAWs span the rectangle in the $x$-direction but not
necessarily in the $y$-direction. For each rectangle 4 
enumerations are done with different restrictions on the
allowed positions of the end-points, e.g., the end-points
may be allowed to lie only on the top border, on the top and/or
bottom border etc. The enumeration of SAWs in a rectangle is
done using the basic transfer matrix algorithm and updating rules
as described above. The only difference is that we have to take
care when creating a new free end that it is allowed under the
restrictions imposed on the end-points. 

If one enumerates anisotropic SAWs in rectangles up to width $W$
it is possible to generate the series correctly to order 
$N_{\rm max} = 4W-1$. Note that this requires the calculation of a
two parameter generating function since variables $u$ and $v$ must be
kept for horizontal and vertical steps, respectively. The generating
functions can be truncated if $m+n > N_{\rm max}$. 

The CEG algorithm requires the calculation of the anisotropic
generating function even though one may ultimately only be
interested in the isotropic SAW counts. However anisotropic series 
can be very useful and most importantly can yield valuable
insights into the analytic properties of the generating function.
In recent papers \cite{EG96,AJG00a} a numerical procedure was given 
(using anisotropic series) that indicates whether or not a given statistical 
mechanics problem is solvable in terms of D-finite functions. A D-finite
function can be expressed as the solution to a linear ordinary differential
equation of finite order with polynomial coefficients.  

\subsection{The pruning algorithm \label{sec:pruning}}

The use of {\em pruning} to obtain more efficient TM algorithms was used 
for SAPs in \cite{JG99}.  We required valid SAPs to span the rectangle in 
{\em both} directions and directly enumerate SAPs of width exactly $W$ and 
length $L$ rather than of width $\leq W$ and length $L$ as done in 
\cite{IGE80e}. At first glance this appears inefficient since we have to 
keep 4 distinct generating functions depending on which borders have been 
touched. However, for SAPs \cite{JG99} it actually leads to an algorithm which 
is both exponentially faster and whose memory requirement is exponentially 
smaller. Experimentally it was found that the computational complexity was 
close to $2^{n/4}$, much better than the $3^{n/4}$ of the original approach.
We have used similar techniques for the enumerations of SAWs carried out 
for this paper.

Pruning, details of which are given in \cite{JG99} for the SAP case, allows 
us to discard most of the possible configurations for large $W$ because they 
only contribute to SAWs of length greater than $N_{\rm max}$, where 
$N_{\rm max}$ is the maximal length to which we choose to carry out our 
calculations. The value of $N_{\rm max}$ is limited by the available 
computational resources, be they CPU time or physical memory. Briefly pruning 
works as follows. Firstly, for each configuration we keep track of the 
current minimum number of steps $N_{\rm cur}$ already inserted to the left
of the boundary line in order to build up that particular configuration. 
Secondly, we calculate the minimum number of additional steps $N_{\rm add}$ 
required to produce a valid SAW. There are three contributions, namely the 
number of steps required to connect the loops and free ends, the number of 
steps needed (if any) to ensure that the SAW touches both the lower and upper 
border, and finally the number of steps needed (if any) to extend at least 
$W$ edges in the length-wise direction (remember we only need rectangles
with $L \geq W$). If the sum $N_{\rm cur}+N_{\rm add} > N_{\rm max}$ we 
can discard the partial generating function for that configuration,
and of course the configuration itself, because it won't make a 
contribution to the SAW count up to the perimeter lengths we are 
trying to obtain. 

There are no principal differences between pruning SAWs and SAPs though
the detailed implementation is more complicated for the SAW case. We found 
it necessary to explicitly write subroutines to handle the three distinct 
cases where the TM configuration contains zero, one and two free ends,
respectively. But in all cases we essentially have to go through all the 
possible ways of completing a SAW in order to find the minimum number of 
steps required. This is a fairly straight forward task though quite time 
consuming. 

Note that the pruning algorithm can be used to generate
either isotropic or anisotropic series. That is, unlike the CEG algorithm,
we need only maintain isotropic generating functions if we are
after isotropic counts for SAWs. But if we wish to do so, say
in order to perform the ``solvability'' check mentioned above
\cite{EG96,AJG00a}, we could calculate anisotropic generating
functions (at the expense of greatly increased memory requirements).

Inspired by Knuth's algorithm for the enumeration of polyominoes 
\cite{Knuth01}, we implemented a couple of further enhancements to our 
SAW algorithm. The first improvement uses a further symmetry of the square 
lattice. When a column has been completed the configuration are symmetric 
under reflection. That is the generating functions for the configurations 
such as, $\{010122030\}$ and $\{030112020\}$, are identical. This symmetry 
also extends to the touching of the upper/lower borders of the rectangle. 
The second improvement is superior memory management. A given boundary line 
configuration only contributes from order $n=N_{\rm cur}+N_{\rm add}$,
so we need only retain the first $N_{\rm max}-n$ terms in the associated 
generating function. In our case the maximum memory consumption occur at 
$W=27$. Here there are approximately $1.12$ billion distinct configurations 
and a total of about  $4$ billion terms in the generating functions. So on 
average there is a little less than 4 terms per configuration. At smaller 
widths there are fewer configurations but more terms per configuration. At 
larger widths both the number of configurations and the number of terms per 
configuration decrease. The important thing to note is that as $N_{\rm max}$ 
is increased the maximal number of terms seems to approach a constant (with 
a value less than 4) times the maximal number of configurations.

\subsection{Computational complexity  \label{sec:comp}}

The time $T(n)$ required to obtain the number of walks of length $n$
grows exponentially with $n$, $T(n) \propto \lambda^n$. For the CEG algorithm 
the complexity can be calculated exactly. Time (and memory) requirements are 
basically proportional to a polynomial (in $n$) times the maximal number of 
configurations, $N_{\rm Conf}$, generated during a calculation. When the 
boundary line is straight and intersects $W+1$ edges we can find the exact 
number of configurations. First look at the situation with no free ends. The 
configurations correspond to Motzkin paths \cite{DV84a} (just map 0 to a 
horizontal step, 1 to a north-east step, and 2 to a south-east step) and the 
number of such paths $M_n$ with $n$ steps is easily derived from the 
generating function
\begin{equation}\label{eq:Motzkin}
M(x) = \sum_{n} M_n x^x = [1-x-(1-2x-3x^2)^{1/2}]/2x^2.
\end{equation}
\noindent
The number of transfer matrix configurations $N_{\rm S}(W)$ in the CEG 
algorithm is simply obtained by inserting 0, 1 or 2
free ends into a Motzkin path and eliminating the path corresponding to
a configuration of all 0's, hence
\begin{equation}\label{eq:Conf}
N_{\rm S}(W) = M_{W+1}+ (W+1)M_W + (W+1)W M_{W-1}/2-1.
\end{equation}
\noindent

When the boundary line has a kink (such as in fig.~\ref{fig:transfer}) 
$N_{\rm Conf}$ is no longer given exactly by (\ref{eq:Conf}). However, it is 
obvious that $N_{\rm S}(W+1) \leq N_{\rm Conf} \leq N_{\rm S}(W)$ so from 
(\ref{eq:Motzkin}) we see that asymptotically $N_{\rm Conf}(W)$ grows like 
$3^W$. Since a calculation using rectangles of widths $\leq W$ yields the 
number of SAW up to $n=4W$ it follows that the complexity of the algorithm 
is $T(n) \propto \lambda^n$ with $\lambda=\sqrt[4]{3}=1.316\ldots$.

\begin{figure}
\begin{center}
\includegraphics[scale=0.5]{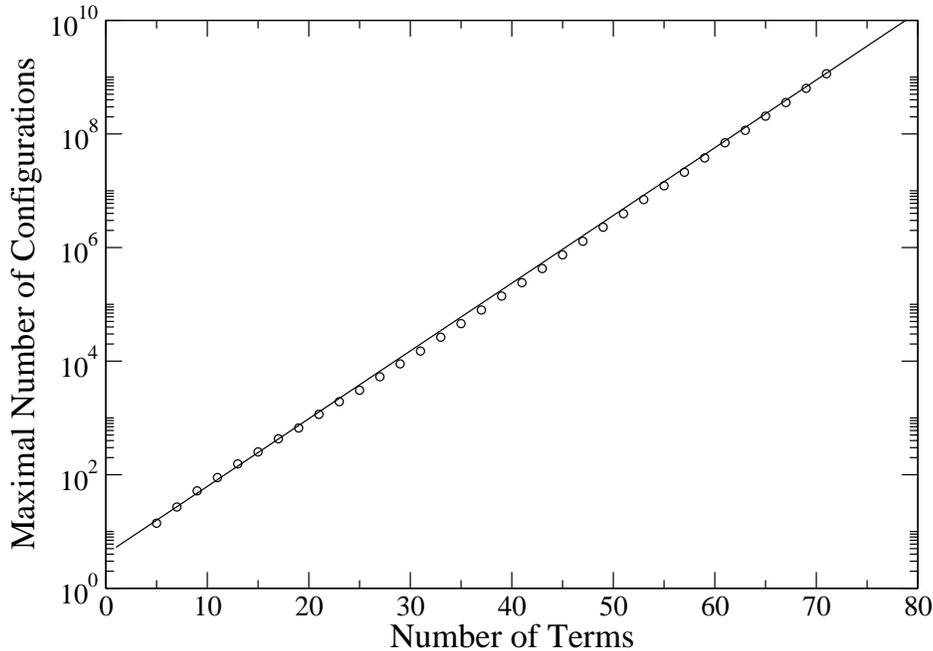}
\end{center}
\caption{\label{fig:memuse}
Lin-log plot of the maximal number of TM configurations from the pruned 
algorithm with increasing $n$.}
\end{figure}

The pruned algorithm is much too difficult to analyse exactly. So all we
can give is a numerical calculation of the growth in the number of configurations
with $n$. That is obtained by just running the SAW algorithm and measuring the maximal 
number of configurations generated for different values of $n$. The resulting graph
is shown in fig.~\ref{fig:memuse}. The straight line, drawn as a guide
to the eye, has slope  $\lambda=\sqrt[4]{3}$ and thus corresponds to the
exponential growth of the CEG algorithm. From this figure it is clear
that the computational complexities of the two algorithms are almost identical.
A closer look at the actual numbers does however reveal that the pruned
algorithm appears to have a slightly higher value of $\lambda_p$. Indeed it 
appears  that increasing $n$ by 8 increases the number of configurations by close 
to a factor of 10 (rather than the 9 expected if $\lambda_p=\sqrt[4]{3}$).
This would mean that for the pruned algorithm 
$\lambda_p \approx \sqrt[8]{10} =1.3335 \ldots$.

The observed value of $\lambda_p$ means that the CEG algorithm is
asymptotically superior to the pruning algorithm, so that for very large
values of $n$ it will be not only be faster but require less
memory as well. However for small $n$ the pruning algorithm is highly
competitive and can in fact use significantly less memory. This
is because the CEG algorithm uses a two parameter generating function
so memory requirements are $\propto n^2 \lambda^n$. For 
the pruning algorithm memory growth is $\propto \lambda_p^n$, rather than
what one may naively have thought $\propto n \lambda_p^n$ (see
comments at the end of the previous section). More concretely, we can mention
that the calculation in  \cite{CG96} of SAWs up to to $n=51$ required
10GB of memory. The pruning algorithm can do the same job using less than
150MB of memory.

\subsection{Parallelization}
\label{SECpara}

The transfer-matrix algorithms used in the calculations of the
finite lattice contributions are eminently suited for parallel
computations. The bulk of the calculations for this paper
were performed on the facilities of the Australian
Partnership for Advanced Computing (APAC). The APAC facility is a HP 
Alpha server cluster with 125 ES45's each with four 1 Ghz chips for a total of 
500 processors in the compute partition. Each server node has at least 
4 Gb of memory. Nodes are interconnected by a low latency high bandwidth 
Quadrics network. 

The most basic concern in any efficient parallel algorithm is
to minimise the communication between processors and ensure that
each processor does the same amount of work and uses the same amount 
of memory. In practice one naturally has to strike some compromise
and accept a certain degree of variation across the processors.

One of the main ways of achieving a good parallel algorithm using 
data decomposition is to try to find an invariant under the
operation of the updating rules. That is we seek to find some property
of the configurations along the boundary line which
does not alter in a single iteration.
The algorithm for the enumeration of SAWs is quite complicated 
since not all possible configurations occur due to pruning
and an update at a given set of edges might change the state of 
an edge far removed, e.g., when two lower loop ends are joined
we have to relabel one of the associated upper loop ends as
a lower loop end in the new configuration.
However, there is still an invariant since any edge not
directly involved in the update cannot change from being 
empty to being occupied and vice versa. That is only the kink edges 
can change their occupation status. This invariant
allows us to parallelise the algorithm in such a way
that we can do the calculation completely independently on each
processor with just two redistributions of the 
data set each time an extra column is added to the lattice. 
We have already used this scheme for SAPs \cite{IJ03a} and 
lattice animals \cite{IJ03b} and refer the interested
reader to these publications for further detail. Our parallelisation
scheme is also very similar to that used by Conway and Guttmann 
\cite{CG96,GC01}.

\subsection{Metric properties}

In a recent paper \cite{IJ00a} we demonstrated that one can use
transfer matrix techniques to calculate the radius of gyration  
of SAPs. Below we show how this work can be extended to calculate
the metric properties of SAWs. 

\subsubsection{Radius of gyrations}

We define the radius of gyration  according to the {\em vertices} of the SAW. 
Note that the number of vertices is one more than the number of steps. The 
radius of gyration of $n+1$ points at positions ${\bf r}_i$ is 

\begin{equation} 
(n+1)^2 \ave{R^2_g}_n = \sum_{i>j} ({\bf r}_i-{\bf r}_j)^2 =
n\sum_i (x_i^2+y_i^2)-2\sum_{i>j}(x_ix_j+y_iy_j).
\end{equation}

This last expression is suitable for a transfer matrix calculation. We 
actually calculate the coefficients of the generating function (\ref{eq:rggf}),
$(n+1)^2 c_n\ave{R^2_g}_n $. In order to do this we have to 
maintain five partial generating functions for each possible boundary 
configuration, namely

\begin{itemize}
\item $C(u)$, the number of (partially completed) SAWs.
\item $X^2_g(u)$, the sum over SAWs of the squared components of the 
distance vectors.
\item $X_g(u)$, the sum of the $x$-component of the distance vectors.
\item $Y_g(u)$, the sum of the $y$-component of the distance vectors.
\item $XY_g(u)$, the sum of the `cross' product of the components of the  
distance vectors, that is, $\sum_{i>j}(x_ix_j+y_iy_j)$.
\end{itemize}

As the boundary line is moved to a new position each target configuration
$S$ might be generated from several sources $S'$ in the previous boundary 
position. The partial generation functions are updated as follows, 
with $(x,y)$ being the coordinates of the newly added vertex:

\begin{eqnarray}\label{eq:rgupdate}
C(u,S) & = & \sum_{S'} u^{n'} C(u,S'), \nonumber \\
X^2_g(u,S) & = & \sum_{S'} u^{n'}[X^2_g(u,S')+\delta_g (x^2+y^2)C(u,S')],\nonumber \\ 
X_g(u,S) & = & \sum_{S'} u^{n'}[X_g(u,S)+ \delta_g xC(u,S')], \\ 
Y_g(u,S) & = & \sum_{S'} u^{n'}[Y_g(u,S)+ \delta_g yC(u,S')], \nonumber \\ 
XY_g(u,S) & = & \sum_{S'}  u^{n'} [XY_g(u,S')+\delta_g xX_g(u,S')
                                  +\delta_g y Y_g(u,S')], \nonumber 
\end{eqnarray}
\noindent
where $n'$ is the number of steps added to the SAW and $\delta_g=0$ if the 
new vertex is empty (has degree 0), $\delta_g=1$ if the new vertex is 
occupied (has degree $>0$).

Finally, when valid SAWs are completed, the partial generating functions are 
added to running totals for each case, and the results for coefficients
in the generating function for the radius of gyration is:

\begin{equation}
(n+1)^2 c_n\ave{R^2_g}_n = n \ave{X^2_g}_n -2\ave{XY_g}_n.
\end{equation}

\subsubsection{End-to-end distance}

The updating rules for the end-to-end distance are very similar to
those for the radius of gyration except that we `count' only the
degree-1 vertices. We again maintain five partial generating functions
for each possible boundary configuration, namely

\begin{itemize}
\item $C(u)$, the number of (partially completed) SAWs.
\item $X^2_e(u)$, the sum over SAWs of the squared components of the 
end-point vectors.
\item $X_e(u)$, the sum of the $x$-component of the end-point vectors.
\item $Y_e(u)$, the sum of the $y$-component of the end-point vectors.
\item $XY_e(u)$, the sum of the `cross' product of the components of the  
end-point vectors.
\end{itemize}

The partial generation functions are updated as described above (\ref{eq:rgupdate})
except that the corresponding quantity $\delta_e=0$ if the new vertex has 
degree 0 or 2, while $\delta_e=1$ if the new vertex has degree 1.

The results for coefficients
in the generating function for the end-to-end distance is:

\begin{equation}
c_n\ave{R^2_e}_n = \ave{X^2_e}_n -2\ave{XY_e}_n.
\end{equation}

\subsubsection{Mean-square monomer distance from end points}

In order to calculate the mean-square distance of a monomer from the end points
we have to introduce an additional  partial generating function

\begin{itemize}
\item $XY_m(u)$, the sum of the `cross' product of the components of the  
end-points and distance vectors.
\end{itemize}
This is updated as follows:

\begin{eqnarray}\label{eq:mdupdate}
XY_m(u,S)  =  \sum_{S'}  u^{n'}[XY_m(u,S') 
          &\!\!\!+\!\!\!& \delta_g xX_e(u,S')+\delta_g y Y_e(u,S') \nonumber \\
          &\!\!\!+\!\!\!& \delta_e xX_g(u,S')+\delta_e y Y_g(u,S')].
\end{eqnarray}
\noindent

The results for the coefficients in the generating function for 
the mean-square monomer distance from end points is :

\begin{equation}
c_n\ave{R^2_e}_m = (n-1)\ave{X^2_e}_n+2\ave{X^2_g}_n-2\ave{XY_m}_n.
\end{equation}

\subsection{Further details}

Finally a few remarks of a more technical nature. The number of contributing 
configurations becomes very sparse in the total set of possible states along 
the boundary line and as is standard in such cases one uses a hash-addressing 
scheme. Since the integer coefficients occurring in the series 
expansion become very large, the calculation was performed using modular 
arithmetic \cite{KnuthACPv2}. This involves performing the calculation modulo 
various integers $p_i$ and then reconstructing the full integer
coefficients at the end. The $p_i$ are called moduli and must be chosen
so they are mutually prime, e.g., none of the $p_i$ have a common divisor.
The Chinese remainder theorem ensures that any integer has a unique 
representation in terms of residues. If the largest absolute values occurring 
in the final expansion is $m$, then we have to use a number of moduli $k$ 
such that $p_1p_2\cdots p_k/2 > m$. Since we are using a heavily loaded 
shared facility  CPU time was more of an immediate limitation than memory. 
So we used the moduli $p_0=2^{62}$ and $p_1=2^{62}-1$, which allowed us to 
represent $p_n$ correctly just using these two moduli.

The calculation of the metric properties require a lot more memory for the 
generating functions, and involves multiplication with quite large integers,
so in this case we used prime numbers of the form $2^{30}-r_i$ for the
moduli $p_i$. Up to 4 primes were needed to represent the coefficients
correctly. 

We were able to extend the series for the square lattice SAW generating 
function from 51 terms to 71 terms using at most 100Gb of memory. The 
calculations requiring the most resource were at widths 24--29. These cases 
were done using 128 processors and took from 16 to 26 hours each.
We also calculated the metric properties of SAWs up to length 59, thus
extending these series from length 32 obtained previously using direct
enumeration. In total the calculations used about 50000 CPU hours.

In table~\ref{tab:cn} we list the number of SAWs from length 52 to 71.
The number of SAWs up to length 51 are tabulated in \cite{GC01} and
\cite{CGJPRS} (this paper also tabulates the metric properties
and several other series). The numbers are also available from
our home page.

\begin{table}[h]
\caption{\label{tab:cn} The number, $c_n$, of embeddings of 
$n$-step SAWs on the square lattice. Only terms for $n>51$ are listed.}
\begin{center}
\small
\begin{tabular}{rrrr} \hline \hline
$n$ & \multicolumn{1}{c}{$c_n$} & $n$ & \multicolumn{1}{c}{$c_n$} \\ \hline 
52  &  37325046962536847970116  &
62  &  646684752476890688940276172   \\
53  &  99121668912462180162908  &
63  &  1715538780705298093042635884   \\
54  &  263090298246050489804708  &
64  &  4549252727304405545665901684   \\
55  &  698501700277581954674604  &
65  &  12066271136346725726547810652   \\
56  &  1853589151789474253830500  &
66  &  31992427160420423715150496804   \\
57  &  4920146075313000860596140  &
67  &  84841788997462209800131419244   \\
58  &  13053884641516572778155044  &
68  &  224916973773967421352838735684   \\
59  &  34642792634590824499672196  &
69  &  596373847126147985434982575724   \\
60  &  91895836025056214634047716  &
70  &  1580784678250571882017480243636   \\
61  &  243828023293849420839513468  &
71  &  4190893020903935054619120005916   \\
\hline \hline
\end{tabular}
\end{center}
\end{table}

\section{Analysis of the series \label{sec:analysis}}

To obtain the singularity structure of the generating functions we used the 
numerical method of differential approximants \cite{AJG89a}. The functions 
have critical points at $u_c$ with exponents as in 
(\ref{eq:sawgf})-(\ref{eq:mdgf}). 
Our main objective is to obtain accurate estimates for the connective 
constant $\mu$ and the critical exponents $\gamma$ and $\nu$. In particular we 
confirm to a very high degree of precision the conjectured exact values 
of the exponents.

Once the exact values of the exponents have been confirmed we turn our
attention to the ``fine structure'' of the asymptotic form of the
coefficients. In particular we are interested in obtaining accurate
estimates for the amplitudes $A$, $C$, $D$ and $E$. We do this by fitting 
the coefficients to the form given by (\ref{eq:sawasymp})-(\ref{eq:md}).
In this case our main aim is to test the validity of the predictions
for the amplitude combinations mentioned in the Introduction.

\subsection{The SAW generating function}

In Table~\ref{tab:anasaw}  we list estimates for the critical point $u_c$  
and exponent $\gamma$ of the series for the square lattice SAW generating 
function. The estimates were obtained by averaging values obtained from 
second and third order differential approximants. For each order $L$ of 
the inhomogeneous polynomial we averaged over those approximants to the 
series which used  at least the first 60 terms of the series.
The error quoted for these estimates reflects the spread (basically one 
standard deviation) among the approximants. Note that these error bounds should
{\em not} be viewed as a measure of the true error as they cannot include
possible systematic sources of error. Based on these estimates we 
conclude that $u_c = 0.379052274(4)$ and $\gamma = 1.343745(15)$.
The estimate for $u_c$ is compatible with the much more accurate
estimate $u_c=0.37905227773(7)$ obtained from the analysis of the SAP 
generating function \cite{IJ03a}. The analysis adds further support to the 
already convincing evidence that the critical exponent 
$\gamma = 43/32=1.34375$ exactly. However, we do observe that
both the central estimates for both $u_c$ and $\gamma$ are systematically
very slightly lower than the expected values.

\begin{table}[h]
\caption{\label{tab:anasaw} Estimates for the critical point
$u_c$ and exponent $\gamma$ obtained from second and third order
differential approximants to the series for square lattice
SAW generating function. $L$ is the order of the inhomogeneous
polynomial.}
\begin{center}
\scriptsize
\begin{tabular}{lllll} \hline \hline
 $L$   &  \multicolumn{2}{c}{Second order DA} & 
       \multicolumn{2}{c}{Third order DA} \\ \hline 
    &  \multicolumn{1}{c}{$u_c$} & \multicolumn{1}{c}{$\gamma$} & 
      \multicolumn{1}{c}{$u_c$} & \multicolumn{1}{c}{$\gamma$} \\ \hline
0   & 0.3790522679(60)  & 1.343735(29) 
    & 0.3790522735(11)  & 1.3437397(18)\\
2   & 0.3790522729(11)  & 1.3437388(23)   
    & 0.3790522752(11)  & 1.3437427(22)\\
4   & 0.3790522720(13)  & 1.3437387(32)   
    & 0.3790522756(27)  & 1.3437438(61)\\
6   & 0.37905227192(81) & 1.3437369(24)   
    & 0.3790522751(27)  & 1.3437429(61)\\
8   & 0.3790522733(15)  & 1.3437395(24)
    & 0.3790522752(27)  & 1.3437434(63)\\
10  & 0.3790522739(30)  & 1.343744(12) 
    & 0.3790522751(22)  & 1.3437430(39)\\
12  & 0.3790522740(19)  & 1.3437404(34)
    & 0.3790522755(63)  & 1.343748(11)\\
14  & 0.3790522738(13)  & 1.3437398(22) 
    & 0.3790522738(25)  & 1.3437406(37)\\
16  & 0.3790522739(12)  & 1.3437403(20)
    & 0.3790522733(39)  & 1.3437408(53)\\
18  & 0.3790522734(14)  & 1.3437398(25) 
    & 0.3790522753(19)  & 1.3437433(41)\\
20  & 0.3790522749(32)  & 1.3437437(87) 
    & 0.3790522755(25)  & 1.3437435(78)\\
\hline \hline
\end{tabular}
\end{center}
\end{table}

\begin{figure}
\begin{center}
\includegraphics{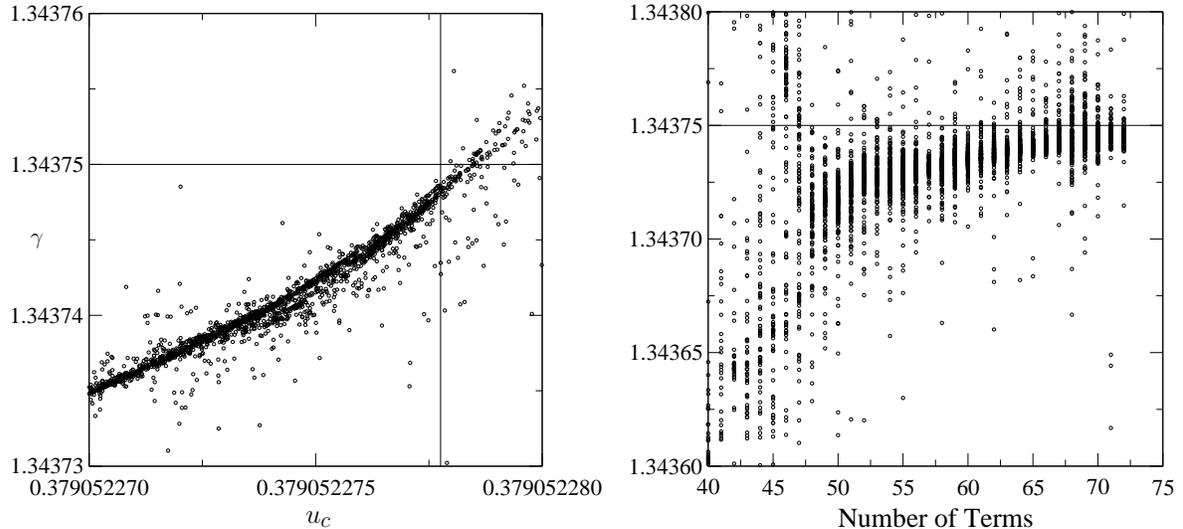}
\end{center}
\caption{\label{fig:sawexp}
Estimates for $\gamma$ vs. $u_c$ (left panel) and $\gamma$ vs. the 
number of terms used by the differential approximant (right panel).}
\end{figure}

When analysing series it is always problematic to get a reliable
error estimate. So in trying to confirm, as we are here, the exact
value of a critical exponent, it is often useful to plot the behaviour
of the estimates against both $u_c$ and the number of terms used
by the differential approximants. In this way it often possible to
gauge more clearly whether or not the high-order approximants have
settled down to the limiting value of the true exponent.
In fig.~\ref{fig:sawexp} we carry out such an analysis. 
Each point in the left panel correspond to  estimates for $u_c$ and 
$\gamma$ obtained from a third order differential approximant. The right 
panel shows the estimates of $\gamma$ but now plotted against the 
number of terms used by the differential approximant. The straight
lines indicate the expected exact value of $\gamma$ and the highly 
accurate estimate of $u_c$ obtained from the analysis of the SAP series.
From the plot in the right panel we can see that the estimates of
$\gamma$ still exhibits a certain up wards drift as the number of
terms in the approximants  increase. So the estimates of $\gamma$ have
not yet settled at their limiting value, but there can be no
doubt that the predicted exact value of $\gamma$ is fully consistent
with the estimates. In the left panel we observe that the 
$(u_c,\gamma)$-estimates fall in a narrow range. Note that
this range does not include the intersection point between
the exact $\gamma$ and the precise $u_c$ estimate. This is
probably a reflection of the lack of `convergence' to
the true limiting values. This view is further supported
by repeating the plot of fig.~\ref{fig:sawexp}, but only using
those approximants using a number of terms is a prescribed
interval, which we choose as 51-55, 56-60, 61-65, and 65-71.
This corresponds to looking at the plots one would have obtained
had the series only been known up to the lengths 55, 60, 65 and
71, respectively. These plots show that as more terms are
included the $(u_c,\gamma)$-estimates move closer and
closer to the expected intersection point. This drift is
again a clear indication that the estimates have not yet
settled at the true limiting values.

\subsection{The metric properties}

We now turn our attention to  the metric properties. The generating functions 
are expected to have a singularity at $u_c$ where the end-to-end distance 
(\ref{eq:eegf}) has exponent $\gamma+2\nu=91/32=2.84375$, the radius
of gyration (\ref{eq:rggf}) has exponent  $\gamma+2\nu+2=155/32=4.84375$,
and the mean square monomer distance from the end points (\ref{eq:mdgf})
has exponent  $\gamma+2\nu+1=123/32=3.84375$. In table~\ref{tab:anametric} 
we list the estimates obtained from a differential approximant analysis
of these series.  In summary we see that applying differential approximants to 
the metric series yields for the end-to-end distance
$u_c=0.37905205(15)$ and $2\nu=2.8434(4)$, the radius of gyration yields 
$u_c=0.379052230(5)$ and $2\nu+2=4.84360(2)$, and the monomer distance yields 
$u_c=0.3790521(1)$ and $2\nu+1=3.84335(15)$. We immediately note that
the exponent estimates are systematically lower that the expected exact values.
Only the end-to-end distance is marginally consistent with
the expected value, while there is a considerable discrepancy between
the radius of gyration estimate and the expected value (similar
though less pronounced for the monomer distance). However, we also
note that the $u_c$ estimates are quite far from the SAP estimate
(in which we have considerable confidence) $u_c=0.37905227773(7)$. So
obviously the metric series are not that well behaved and might
have large corrections to scaling.

\begin{table}
\caption{\label{tab:anametric} Estimates for the critical point
$u_c$ and critical exponents obtained from second and third order
differential approximants to the series for the end-to-end distance, 
radius of gyration, and monomer distance from the end
point.}
\begin{center}
\scriptsize
\begin{tabular}{lllll} \hline \hline
         \multicolumn{5}{c}{$\RGf_e (u)$}  \\ \hline
 $L$   &  \multicolumn{2}{c}{Second order DA} & 
       \multicolumn{2}{c}{Third order DA} \\ \hline 
    &  \multicolumn{1}{c}{$u_c$} & \multicolumn{1}{c}{$\gamma+2\nu$} & 
      \multicolumn{1}{c}{$u_c$} & \multicolumn{1}{c}{$\gamma+2\nu$} \\ \hline
0   & 0.379052003(90)  & 2.84324(62)
    & 0.379052101(69)  & 2.84333(19)\\
2   & 0.379051985(57)  & 2.84301(96)
    & 0.379052116(58)  & 2.84336(11)\\
4   & 0.379052046(81)  & 2.84345(18)
    & 0.379052113(75)  & 2.84336(10)\\
6   & 0.379052034(80)  & 2.84329(39)
    & 0.379052119(86)  & 2.84334(19)\\
8   & 0.379052054(69)  & 2.8430(19)
    & 0.379052115(75)  & 2.84337(33)\\
10  & 0.379052035(67)  & 2.84329(23)
    & 0.379052138(66)  & 2.84338(11)\\ 
\hline
         \multicolumn{5}{c}{$\RGf_g (u)$}  \\ \hline
 $L$   &  \multicolumn{2}{c}{Second order DA} & 
       \multicolumn{2}{c}{Third order DA} \\ \hline 
3    &  \multicolumn{1}{c}{$u_c$} & \multicolumn{1}{c}{$\gamma+2\nu+2$} & 
      \multicolumn{1}{c}{$u_c$} & \multicolumn{1}{c}{$\gamma+2\nu+2$} \\ \hline
0   & 0.3790522317(19)  & 4.8436019(22) 
    & 0.3790522289(11)  & 4.8435986(13)\\
2   & 0.3790522317(26)  & 4.8436017(29)
    & 0.3790522295(10)  & 4.8435992(11)\\
4   & 0.3790522324(41)  & 4.8436024(45)
    & 0.3790522289(22)  & 4.8435986(23)\\
6   & 0.3790522290(94)  & 4.843598(11)
    & 0.3790522284(23)  & 4.8435980(26)\\
8   & 0.379052225(15)   & 4.843595(16)
    & 0.3790522294(43)  & 4.8435992(48)\\
10  & 0.3790522282(21)  & 4.8435978(24)
    & 0.3790522301(20)  & 4.8436000(24)\\ 
\hline
         \multicolumn{5}{c}{$\RGf_m (u)$}  \\ \hline
 $L$   &  \multicolumn{2}{c}{Second order DA} & 
       \multicolumn{2}{c}{Third order DA} \\ \hline 
    &  \multicolumn{1}{c}{$u_c$} & \multicolumn{1}{c}{$\gamma+2\nu+1$} & 
      \multicolumn{1}{c}{$u_c$} & \multicolumn{1}{c}{$\gamma+2\nu+1$} \\ \hline
0   & 0.379052045(58)  & 3.84321(11)
    & 0.379052131(39)  & 3.843345(61)\\
2   & 0.379052056(37)  & 3.843256(63)
    & 0.379052118(57)  & 3.843327(93)\\
4   & 0.379052044(70)  & 3.84323(10)
    & 0.379052107(68)  & 3.84332(10)\\
6   & 0.379052050(73)  & 3.84322(10)
    & 0.379052088(51)  & 3.843281(96)\\
8   & 0.379052081(99)  & 3.84329(17)
    & 0.379052081(52)  & 3.843274(95)\\
10  & 0.379052069(95)  & 3.84326(17)
    & 0.379052135(55)  & 3.843370(86)\\ 
\hline \hline
\end{tabular}
\end{center}
\end{table}

\begin{figure}
\begin{center}
\includegraphics[scale=0.95]{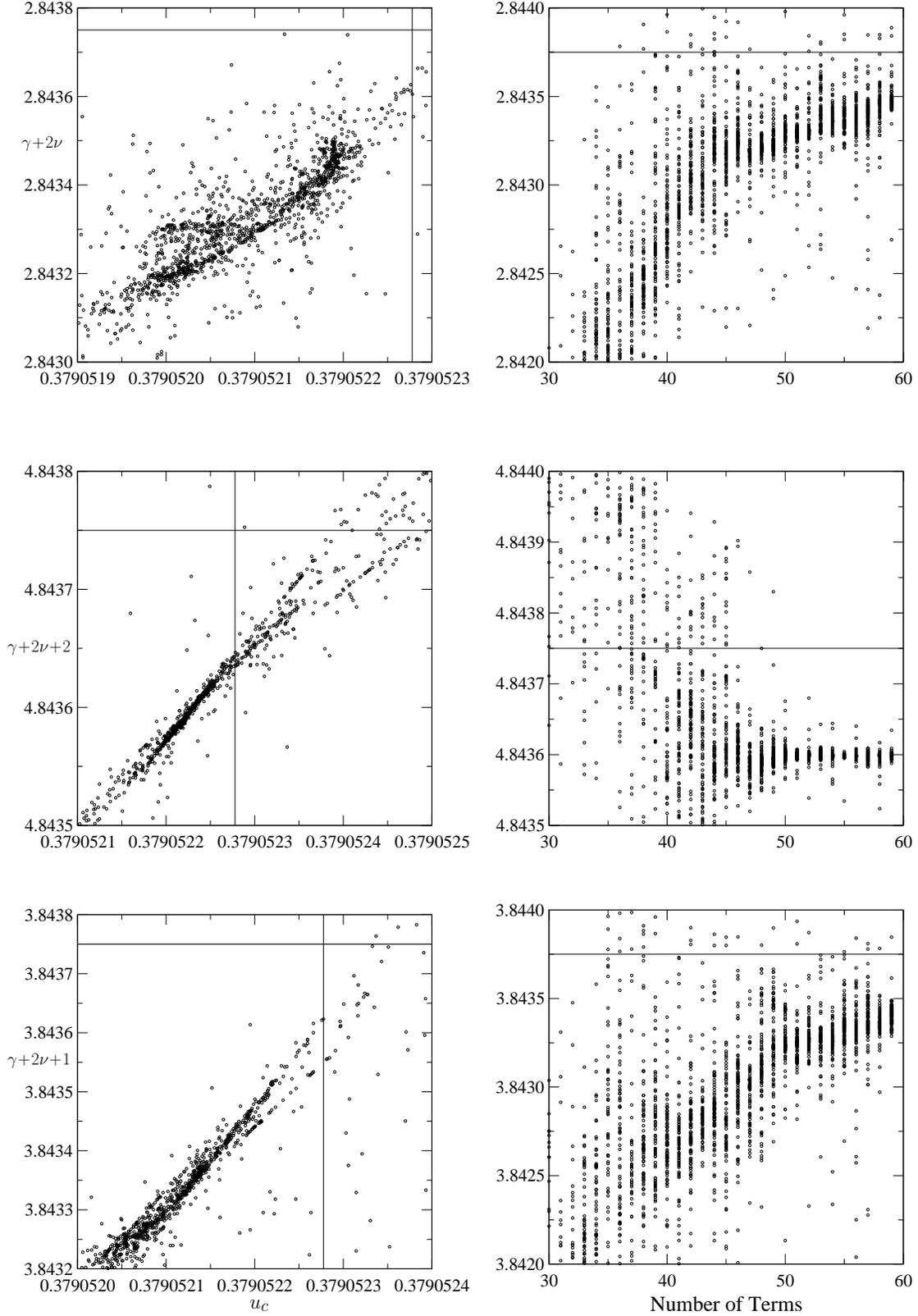}
\end{center}
\caption{\label{fig:metric}
Plots of estimates for the critical exponents vs. the critical
point (left panels) and critical exponents vs. number of terms (right panels)
obtained form third order differential approximants to the generating functions
$\RGf_e (u)$ (top panels), $\RGf_g (u)$ (middle panels) and $\RGf_m (u)$ 
(bottom panels).}
\end{figure}

As for the SAW generating function, we find it useful to plot the
estimates for the critical exponents vs. $u_c$ and the number
of terms. This we have done in fig.~\ref{fig:metric}. Clearly
the estimates from the end-to-end distance have not yet converged
and it is quite possible that the exponent estimates will eventually converge to
the expected value (see top left panel). Also in the top right panel it 
is quite possible that the estimates will approach the point given by 
the intersection of the exact exponent value and the precise $u_c$ value. 
The behaviour of the estimates for radius of gyration and monomer
distance series are far more unsettling. In particular, the exponent
estimates from the radius of gyration series appears well converged to 
a value $4.84360$ with a narrow spread which clearly does not
include the expected exact value $4.84375$ and in the plot
of the  exponent vs. $u_c$ (middle left panel) the estimates 
are quite far from the expected intersection. Similar remarks
hold for the monomer distance (bottom panels) though convergence
and discrepancy with expected values is less pronounced. 
Furthermore, the behaviour of the radius of gyration series is
very different to the other series. In particular we note that
in the plots of the exponents vs. the number of terms (left panels)
the estimates from the end-to-end and monomer distance
seems to increase monotonically to wards the expected value, while
the estimates from the radius of gyration starts out above the
expected value, then cross the expected value before apparently
settling  below the expected value. This behaviour is quite
puzzling. Let us just note that if we look at the similar plots for
the SAW generating function (fig.~\ref{fig:sawexp}) it is clear
that convergence has not been achieved at $n=59$ and not yet even at
$n=71$. It would therefore be surprising if we should not see a
further drift in the exponent estimates for the metric properties
with longer series.

\begin{figure}[h]
\begin{center}
\includegraphics{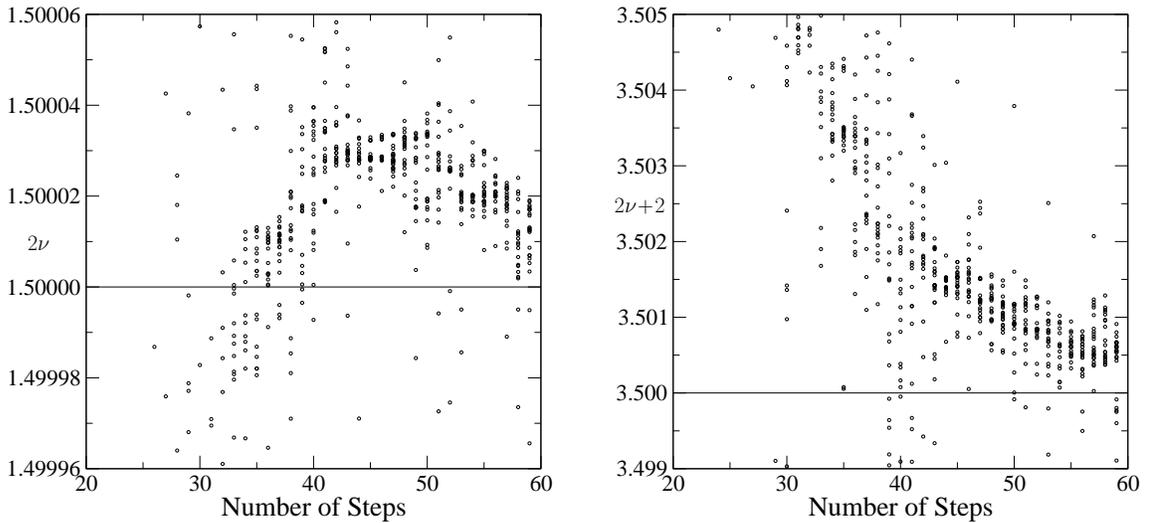}
\end{center}
\caption{\label{fig:metric_div}
Plots of estimates of the critical exponents versus the number of terms 
obtained form third order differential approximants to the functions
$\RGf_e (u)/(u\CGf(u))$ (left panel) and  
$\RGf_g (u)/(u\CGf(u))$ (right panel).}
\end{figure}

Fortunately, we have the possibility of analysing other series
involving the metric properties. We can look
directly at the generating function with coefficients $\ave{R^2_e}_n$,
and so on. While this has the advantage that we know that these
series have a critical point at $u_c=1$ it turns out that the 
estimates of the critical exponents, $-2\nu-1$ in all cases,
behave in exactly the same manner as the series for the
original generating functions. A second, and as we shall see
much more successful approach, is to take the original series
and divide them by the SAW generating function. That is we study
the series $\RGf_e(u)/(u*\CGf(u)) \propto (u-u_c)^{-2\nu}$, and so on,
where again $u_c=1/\mu$. We won't go through all the details here.
Suffice to say that applying differential approximants to the
resulting series yields for the end-to-end distance series
$u_c=0.37905229(2)$ and $2\nu=1.50002(3)$, the radius of gyration series yields 
$u_c=0.3790526(4)$ and $2\nu+2=3.5006(4)$, and the monomer distance series yields 
$u_c=0.3790524(2)$ and $2\nu+1=2.5002(4)$. This clearly confirms
that $\nu=3/4$, exactly. We notice that the error estimates of
the modified series is quite different to the original series.
The original end-to-end series has the largest error estimate of the
three while the modified  end-to-end series has the smallest error estimate.
The opposite happens for the radius of gyration series.  
The quite different behaviour of these
series, as compared to the original ones, is probably even more
clearly illustrated in fig.~\ref{fig:metric_div} where we have
plotted the exponent estimates vs. the number of terms for the
modified end-to-end and radius of gyration series. Clearly, the
estimates from both of these series appear not yet to have settled at 
their limiting values, but it would seem that they are converging
to wards the expected exponent values. Note the very different behaviour
of the original and modified radius of gyration series.
So obviously dividing by the SAW generating function has a dramatic
effect on the metric series. We can only guess that this procedure
leads to modifications in the correction-to-scaling behaviour
thus altering dramatically the convergence properties of the series.

\subsection{Non-physical singularities}

The generating functions have singularities on the
negative axis at $u_-=-1/\mu=-x_c$. The exponents at $u_-$ are compatible
with simple exact values. For the SAW generating function the
exponent is $1/2$, for the end-to-end generating function (\ref{eq:eegf})
the exponent is also $1/2$, for the radius of gyration generating function 
(\ref{eq:eegf}) the exponent is $-3$, while for the monomer distance 
generating function (\ref{eq:mdgf}) the exponent is $-2$.

\subsection{Correction-to-scaling exponent}

The correction-to-scaling exponent for SAWs is exhaustively studied in a recent paper 
\cite{CGJPRS} using series analysis and Monte Carlo simulations. 
In particular we performed a very careful and detailed analysis of the 
of the 59 step series for the square lattice SAW counts and metric properties 
and a 40 step series for the triangular lattice. In that study of the SAW 
correction-to-scaling exponents, a consistent picture emerged. We presented 
compelling evidence that the first non-analytic correction term in the 
generating function for SAWs and SAPs is $\Delta_1 = 3/2$, as predicted by 
Nienhuis \cite{Nienhuis82a,Nienhuis84a}. We found no evidence for the 
presence of an exponent $\Delta_1 = 11/16$ in SAWs and SAPs on the square 
and triangular lattices, as proposed by Saleur \cite{Saleur87a}.

Our method of analysis, both here and in \cite{CGJPRS}, is based on direct fitting
to the expected asymptotic form. Obviously (\ref{eq:sawasymp}) only
gives the leading term in the asymptotic expansion. We have to add in 
both analytic and non-analytic corrections to scaling. Furthermore, we 
have to take account of the presence of the singularity at $u_-=-1/\mu$. We thus 
expect $c_n$ to have an asymptotic expansion of the form 
\begin{equation}\label{eq:cnasymp}
c_n \sim \mu^n n^{\gamma-1}[a_0 + \sum_{i=1}^k \frac{a_i}{n^{\Delta_i}}] +
(-\mu)^n n^{\alpha - 2}[b_0 + \sum_{i=1}^m \frac{b_i}{n^{\Gamma_i}}],
\end{equation}
where $\alpha$ is the critical exponent occurring in the polygon
generating function. 

We estimate the coefficients $a_i$ and $b_i$, by inserting
the estimated value of $\mu,$ the exact values of $\gamma$ and $\alpha$,
and assumed values of $\Delta_i$ and $\Gamma_i$. The coefficients can then be
fitted to the assumed asymptotic form by solving a system of linear equations.
By steadily increasing the number of series coefficients, many estimates
for the $\{a_i\}$ and $\{b_i\}$ are found.  Provided the different estimates
are consistent over series of different lengths we assume that they provide
an acceptably accurate estimate of the actual asymptotic coefficients.

A noteworthy feature of the method is that, if a blatantly false exponent
is given as input (for example, specifying $\Delta_1 = 1/2$ for the 
two-dimensional SAW), the sequence of amplitude estimates for the term 
corresponding to that exponent will converge rapidly to zero, giving a 
very strong signal that the exponent in question is absent. It was
this  feature which was used to rule out $\Delta_1 = 11/16$.

Note that one would expect a whole sequence $\Delta_j>\Delta_1$ of 
non-analytic corrections to scaling, as well as so-called mixing
terms involving the exponents $\gamma$ and $\alpha$ (see \cite{CGJPRS} for
details). The first expected mixing term would give a contribution  
$n^{-59/32}$ \cite{CGJPRS}. However, in practice this is indistinguishable from 
the $n^{-2}$ term. Higher order corrections can likewise not be detected since
they make contributions well beyond the range of reasonable extrapolation.

\subsection{Amplitudes}

In our paper \cite{CGJPRS} we also obtained accurate amplitude estimates. 
Here we shall therefore only briefly review these results then report on 
the slightly improved estimates for the amplitudes of the SAW counts based 
on the extended 71 term series.

Given the value for the non-analytic correction-to-scaling exponent we 
more concretely choose to fit fit to the form used in \cite{CGJPRS}:
\begin{eqnarray}
c_n & \sim & \mu^n n^{11/32}[a_0 + a_1/n + a_2/n^{3/2}  
             + a_3/n^2 + a_4/n^{5/2} + \cdots] \nonumber \\
   &+ & (-1)^n \mu^n n^{-3/2}[b_0 + b_1/n + b_2/n^2 + b_3/n^3 + \cdots].
\label{eq:cnasympexpl}
\end{eqnarray}
Fitting to this form we found \cite{CGJPRS}, using the 59 term SAW counts, 
$a_0=A=1.1770425(5)$ as well as reasonably accurate estimates for 
$a_1$--$a_3$ and $b_0$--$b_2$. For the metric properties we found
$C=0.77124(8)$, $D=0.108227(58)(5)$ and $E=0.33913(14)$.
A similar analysis of the triangular lattice data
yielded $A=1.183966(2)$, $C=0.71176(66)$, $D=0.09987(5)$ and $E=0.3130(5)$.

\begin{table}[h]
\caption{\label{tab:amplratio} Estimates of universal
amplitude combinations on some two-dimensional lattices.}
\begin{center}
\begin{tabular}{lllll} \hline \hline
Lattice &  $D/C$ & $E/C$ & $BC/\sigma a_0$ & $ F$ \\ \hline
Square \cite{CGJPRS,IJ03a}     & 0.140299(6) & 0.439647(6) & 0.21683(4) & -0.000024(28) \\
Triangular \cite{CGJPRS,IJ04d} & 0.140296(6) & 0.439649(9) & 0.2169(2) & -0.000036(34)\\
Honeycomb \cite{Lin00}   & 0.1403(1) &   0.4397(2) & 0.2170(3) & -0.00013(67) \\
Kagom\'e \cite{Lin95a,Lin99b}   & 0.140(1) & 0.440(1) & 0.2144(25) & -0.0015(47) \\
\hline \hline
\end{tabular}
\end{center}
\end{table}

The ratios $D/C$ and $E/C$ were also estimated by direct extrapolation
of the relevant quotient sequence, using the following method \cite{OPBG}:
Given a sequence $\{a_n\}$ defined for $n \ge 1$,
assumed to converge to a limit $a_{\infty}$
with corrections of the form
$a_n \sim a_{\infty}(1 + b/n + \ldots)$,
we first construct a new sequence $\{h_n\}$
defined by $h_n = \prod_{m=1}^n a_m$.
The generating function
$\sum h_n x^n \sim (1 - a_{\infty} x)^{-(1+b)}$.
Estimates for $a_{\infty}$
and the parameter $b$ can then be obtained from differential approximants.
In this way, we obtained the estimates \cite{CGJPRS},
$D/C = 0.140299(6)$ and $E/C = 0.439647(6)$ for the square lattice and
$D/C = 0.140296(6)$ and $E/C = 0.439649(9)$ for the triangular lattice.

The amplitude estimates leads to a high precision confirmation of the
CSCPS relation (\ref{eq:CSCPS}) $F=0.000024(25)$.

In Table~\ref{tab:amplratio} we have listed the estimates of
various universal amplitude combinations, obtained from the work
reported in this paper and elsewhere. As can be seen the estimates
for all lattices are in perfect agreement strongly confirming the
universality of the various combinations. 

\begin{figure}
\begin{center}
\includegraphics{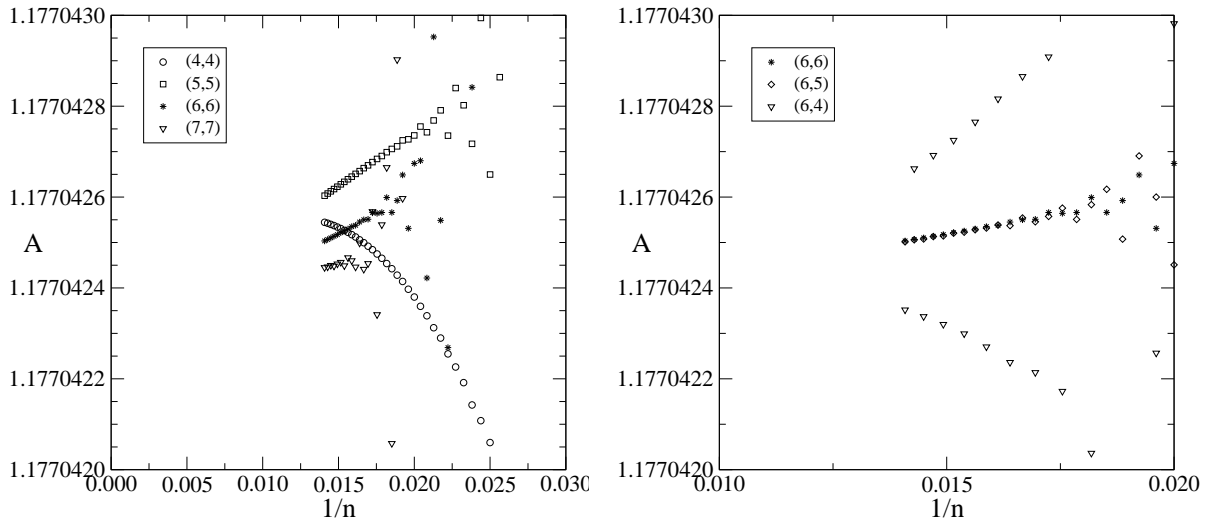}
\end{center}
\caption{\label{fig:ampl}
Plots of estimates for the leading amplitude $A$ from various fits of $
c_n$ to the assumed asymptotic expansion (\ref{eq:cnasymp}).}
\end{figure}

Finally we turn to the estimation of the amplitude $A$ using
the new extended 71 term series. As in previous work \cite{IJ00a,IJ03a}
we find it very useful to plot the amplitude estimates vs. $1/n$ where
$c_n$ is the last coefficient used by the fit. In fig.~\ref{fig:ampl} we
plot the estimates for the leading amplitude $A$ from various 
fits. The legend numbers $(k,m)$ indicates the number of terms used
in the fit by each part of the asymptotic expansion (\ref{eq:cnasymp}),
using the exponents given in the  explicit form  (\ref{eq:cnasympexpl}).
From the left panel we see a consistent trend emerging. As the number
of terms used in $(k,k)$-fits is increased we see that the estimates
settle down and that fits using more terms display less curvature.
We take this as a clear indication that the fitting procedure is robust
and that the assumed asymptotic expansion is correct. The fits using
5, 6, and 7 terms each can clearly be extrapolated to a value
$A=1.770423(1)$. In the right panel we plot amplitude estimates
from $(k,m)$-fits with $k=6$ and $m=6$, 5 and 4. We do this merely
to point out that clearly only $(k,m)$-fits with $m$ close to $k$ are 
reliable. The $(6,4)$-fit displays a pronounced oscillatory behaviour.

\section{Summary and conclusion}

We have presented a new algorithm for the enumeration of self-avoiding
walks. Numerical data show that it has computational complexity
only slightly worse than the Conway-Enting-Guttmann algorithm \cite{CEG93}.
This means that the CEG algorithm will be superior for enumerating
very long SAWs. However, the new algorithm uses much less memory at shorter
lengths and remains competitive at lengths attainable at present and in the 
foreseeable future.
Furthermore the new algorithm can be used to calculate metric properties
such as the end-to-end distance, radius of gyration, and average distance 
of monomers from the end points. We have used the algorithm to extend the 
series for the number of SAWs on the square lattice up to 71 steps and 
calculate the metric properties of SAWs up to 59 steps.

Analysis of the series yielded estimates of the critical exponents
$\gamma$ and $\nu$ which confirmed to a high degree of accuracy
the predicted exact values $\gamma=43/32$ and $\nu=3/4$.
We reported on results from a comprehensive analysis \cite{CGJPRS} of the
series providing very firm and convincing evidence that the
leading non-analytic correction to scaling is $\Delta_1=3/2$,
as well as giving accurate estimates for the critical amplitudes.
The amplitude estimates led to a high precision confirmation of the
CSCPS relation (\ref{eq:CSCPS}) $F=0$.

\section{Acknowledgments}

It is a pleasure to thank Tony Guttmann for a careful reading of the
manuscript and many helpful suggestions.
The calculations presented in this paper would not have been possible
without a generous grant of computer time on the server cluster of the
Australian Partnership for Advanced Computing (APAC). We also used
the computational resources of the Victorian Partnership for Advanced 
Computing (VPAC). We gratefully acknowledge financial support from 
the Australian Research Council.

\end{document}